\documentclass[sigconf,authorversion=true, nonacm=true]{acmart}
\usepackage{caption}
\usepackage[skip=0cm,list=true]{subcaption}
\usepackage{color,soul}
\usepackage{braket}

\AtBeginDocument{%
  \providecommand\BibTeX{{%
    \normalfont B\kern-0.5em{\scshape i\kern-0.25em b}\kern-0.8em\TeX}}}

\copyrightyear{2022}
\acmYear{2022}
\setcopyright{acmcopyright}\acmConference[NANOCOM '22]{The Ninth Annual ACM International Conference on Nanoscale Computing and Communication}{October 5--7, 2022}{Barcelona, Spain}
\acmBooktitle{The Ninth Annual ACM International Conference on Nanoscale Computing and Communication (NANOCOM '22), October 5--7, 2022, Barcelona, Spain}
\acmPrice{15.00}
\acmDOI{10.1145/3558583.3558846}
\acmISBN{978-1-4503-9867-1/22/10}

\begin{document}

\title[Characterizing Qubit Traffic of a Quantum Intranet aiming at Modular Quantum Computers]{Characterizing the Spatio-Temporal Qubit Traffic of a Quantum Intranet Aiming at Modular Quantum Computer Architectures}





\author{Santiago Rodrigo}
\email{srodrigo@ac.upc.edu}
\orcid{0000-0001-8843-5276}
\affiliation{%
  \institution{Universitat Polit\`ecnica de Catalunya}
  \city{Barcelona}
  \country{Spain}
}

\author{Domenico Span\`{o}}
\email{spndnc98p02c710a@studenti.unirc.it}
\affiliation{%
  \institution{U. Mediterranea di Reggio Calabria}
  \city{Reggio Calabria}
  \country{Italy}
}

\author{Medina Bandic}
\email{m.bandic@tudelft.nl}
\orcid{0000-0003-4670-0988}
\affiliation{%
\institution{Delft University of Technology}
  \city{Delft}
  \country{The Netherlands}
}
  
\author{Sergi Abadal}
\email{abadal@ac.upc.edu}
\orcid{0000-0003-0941-0260}
\affiliation{%
  \institution{Universitat Polit\`ecnica de Catalunya}
  \city{Barcelona}
  \country{Spain}
}

\author{Hans van Someren}
\email{j.vansomeren-1@tudelft.nl}
\orcid{0000-0003-4763-6455}
\affiliation{%
\institution{Delft University of Technology}
  \city{Delft}
  \country{The Netherlands}
}

\author{Anabel Ovide}
\email{anabel.ovide@gmail.com}
\affiliation{%
  \institution{University of Tartu}
  \city{Tartu}
  \country{Estonia}
}

\author{Sebastian Feld}
\email{s.feld@tudelft.nl}
\orcid{0000-0003-2782-1469}
\affiliation{%
  \institution{Delft University of Technology}
  \city{Delft}
  \country{The Netherlands}
}

\author{Carmen G. Almud\'{e}ver}
\email{cargara2@disca.upv.es}
\orcid{0000-0002-3800-2357}
\affiliation{%
  \institution{Universitat Polit\`ecnica de Val\`encia}
  \city{Valencia}
  \country{Spain}
}

\author{Eduard Alarc\'on}
\email{eduard.alarcon@upc.edu}
\orcid{0000-0001-7663-7153}
\affiliation{%
  \institution{Universitat Polit\`ecnica de Catalunya}
  \city{Barcelona}
  \country{Spain}
}

\renewcommand{\shortauthors}{S. Rodrigo et al.}

\begin{abstract}
Quantum many-core processors are envisioned as the ultimate solution for the scalability of quantum computers. Based upon Noisy Intermediate-Scale Quantum (NISQ) chips interconnected in a sort of \emph{quantum intranet}, they enable large algorithms to be executed on current and close future technology. In order to optimize such architectures, it is crucial to develop tools that allow specific design space explorations. To this aim, in this paper we present a technique to perform a spatio-temporal characterization of quantum circuits running in multi-chip quantum computers. Specifically, we focus on the analysis of the qubit traffic resulting from operations that involve qubits residing in different cores, and hence quantum communication across chips, while also giving importance to the amount of intra-core operations that occur in between those communications. Using specific multi-core performance metrics and a complete set of benchmarks, our analysis showcases the opportunities that the proposed approach may provide to guide the design of multi-core quantum computers and their interconnects.
\end{abstract}

\ccsdesc[500]{Computer systems organization~Quantum computing}
\ccsdesc[500]{Computer systems organization~Multicore architectures}
\ccsdesc[500]{Networks~Network performance analysis}
\ccsdesc[300]{Networks~Network simulations}
\ccsdesc[300]{Networks~Network on chip}
\keywords{Quantum Computing, Multicore Quantum Architectures, Traffic Characterization, Network Performance Analysis, Network-on-Chip}


\maketitle

\section{Introduction}

\begin{figure*}
\centering
\begin{subfigure}[t]{0.63\textwidth}
  \centering
  \includegraphics[width=\linewidth]{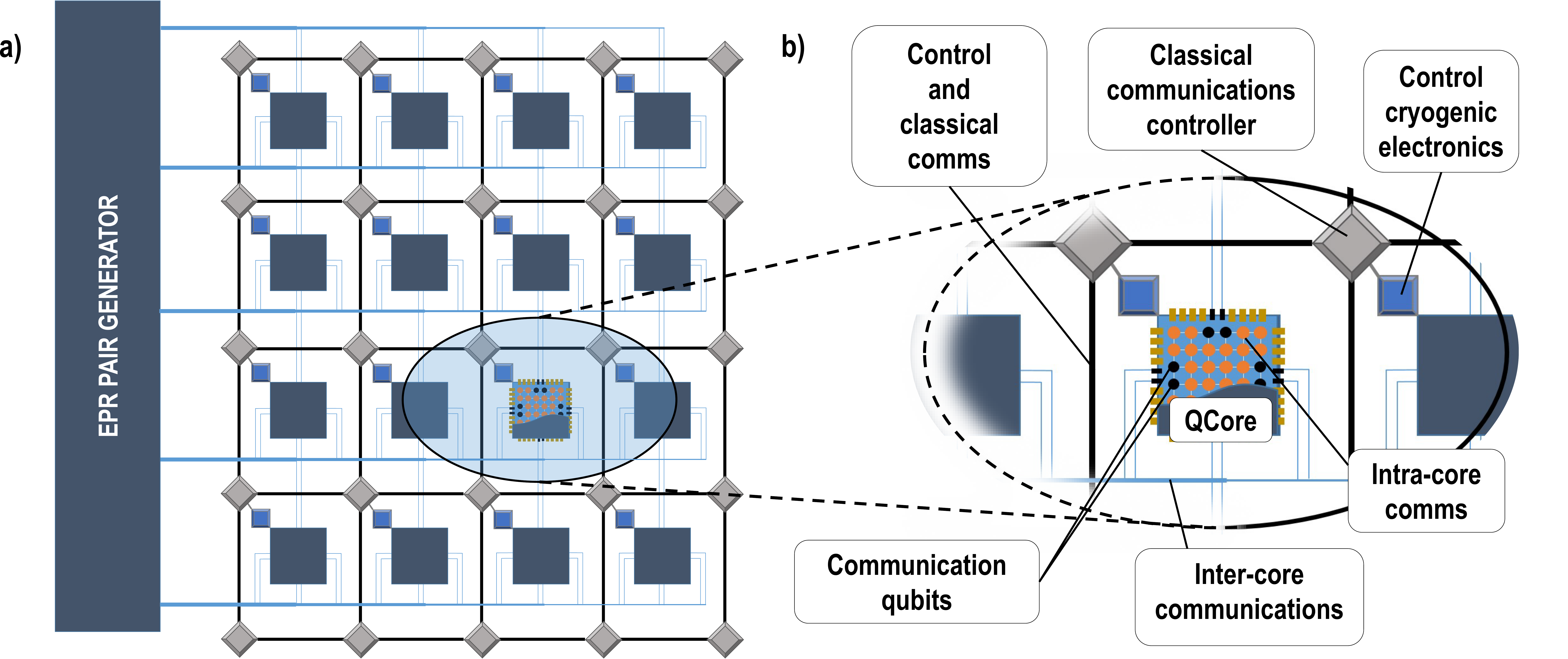}
  \vspace{-0.3cm}
  \vspace{-0.1cm}
  \label{fig::multi-core_computer_sub1}
\end{subfigure}%
\begin{subfigure}[t]{0.24\textwidth}
  \centering
  \includegraphics[width=\linewidth]{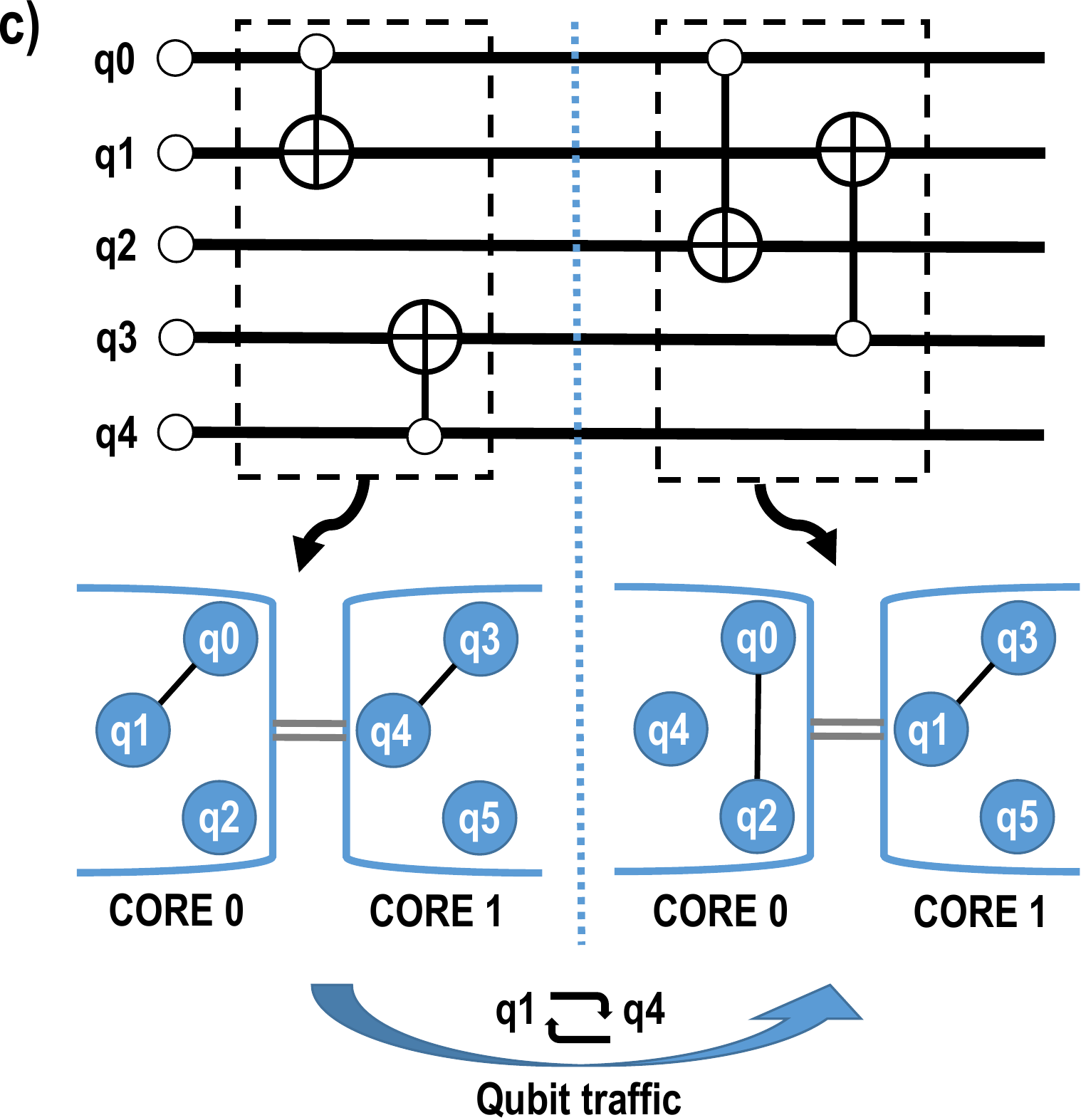}
  \vspace{-0.3cm}
  \vspace{-0.1cm}
  \label{fig::multi-core_computer_sub2}
\end{subfigure}
\vspace{-0.3cm}
\caption{\textbf{Multi-core quantum computing.} \textbf{a)} 2D diagram of a multi-chip architecture. The classical network also depicted completes the networking infrastructure. \textbf{b)} Enumeration of the components, including intra- and inter-core communications. \textbf{c)} Qubit traffic generated by internal core topology and inter-core operations.}
\vspace{-0.1cm}
\label{fig::multi-core_computer}
\end{figure*}

Quantum Computing theoretical capabilities are beyond doubt, and almost fifty years of research have prepared the stage for what some call the Second Quantum Revolution \cite{dowling2003quantum}: an unprecedented \textit{quantum} leap in key areas such as cryptography, biochemistry, big data analysis, or artificial intelligence\cite{martonosi2019next}. However, although during the last decade the advances in quantum computer prototypes are impressive, they are still struggling with integrating tens to hundreds of qubits \cite{IBM2019qcomp53qubits,arute2019quantum,chow2021ibm}, far away from the amount needed for fully-fledged systems. Hard engineering challenges related with the unstable nature of quantum states (quantum decoherence, need for per-qubit control, and others) are still today hindering the scalability of this technology~\cite{almudever2017engineering}.

As an alternative way to ongoing research on better qubit isolation and control leading to the integration of more qubits in the same chip, in the last years several groups have proposed putting together currently available small-sized computing nodes and making them work coordinately. At first they were theoretical approaches aiming at widening the research frontier, presenting two different approaches: distributed quantum computing, related to the development of the Quantum Internet \cite{cacciapuoti2019quantum,van2007communication,caleffi2018quantum}, and short-range multi-core quantum computers on-a-chip \cite{monroe2014large,brown2016co,vandersypen2017interfacing,isailovic2006interconnection}. While large-scale Quantum Internet seems very promising, important challenges (e.g. interfacing of qubits at cryogenic temperatures and communicating photons at room temperature) currently prevent its full development and the implementation of distributed quantum computing platforms. On the other hand, latest proposals on short-range communications linking existing NISQ processors include practical analysis and simulations on this type of architectures \cite{rodrigo2021modelling, rodrigo2021scaling}. Very relevantly, IBM has very recently released their updated quantum roadmap with multi-chip communication and teamwork computing as its keystone \cite{gambetta2022expanding}.

However, interconnecting quantum processors comes with its own share of hard challenges. Quantum data cannot be copied and latencies are crucial, as quantum decoherence steadily corrupts qubits. This, together with the need for computing with qubits in-place, calls for an entangled design between communications and computation in multi-core quantum computer architectures~\cite{rodrigo2021double}. Such an approach should strive to minimize the communications overhead and optimize the computing efficiency when distributing qubits among different processors, since moving qubits around is expensive and affects directly the reliability of the whole computation. To that end, however, techniques for the analysis of the quantum communication traffic on these architectures is required. 


In this paper, we aim to bridge this gap by providing a spatio-temporal analysis of qubit traffic, which allows to assess the impact of the quantum algorithm, mapping process and architecture on the actual execution. Based on this tool, we present some first results on traffic characterization of multi-core quantum architectures, which suggest that conscious design and optimization of compilers for multi-core quantum architectures could solve current issues in the amount and distribution (over time and space) of inter-core quantum data transfers.

In order to set a common ground for the discussion, we summarize in Section 2 the key concepts on quantum computing and communication. The use of traffic analysis on classical multi-core computing is reviewed in Section 3, discussing how this should be adapted into quantum computing and highlighting the need for a deeper understanding of multi-core quantum computers' inner workings in order to optimize these architectures. Section 4 is devoted to describing our qubit traffic analysis procedure, and in Section 5 we present some first results based on that tool, before concluding the paper.

\section{Background}
\label{sec:background}

\begin{figure*}
\centering
\begin{subfigure}[t]{0.3\textwidth}
  \centering
  \includegraphics[width=\linewidth]{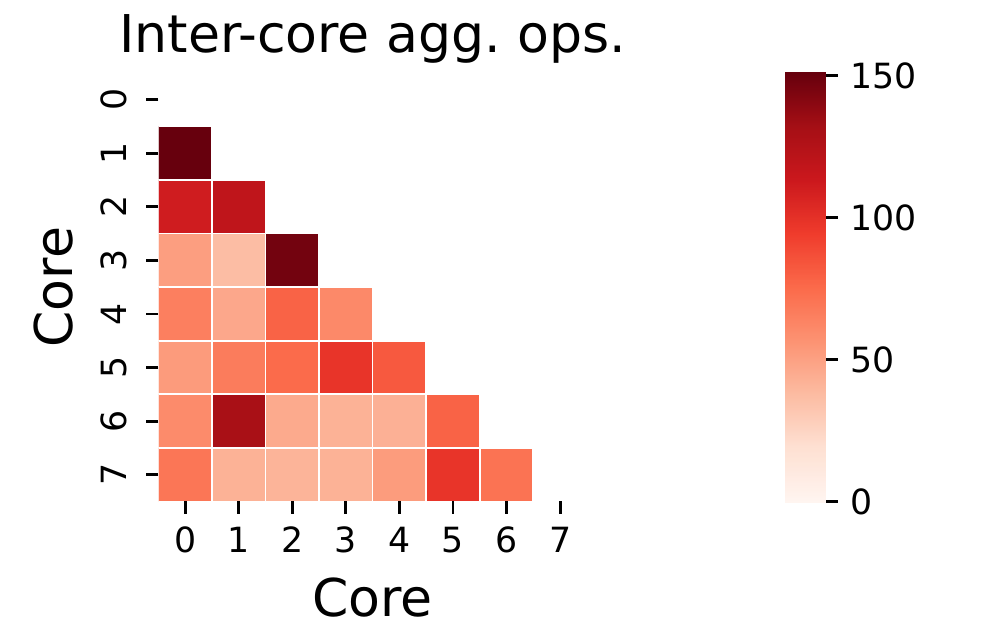}
  \vspace{-0.3cm}
  \caption{}
  \vspace{-0.1cm}
  \label{fig::traffic_heatmap_sub1}
\end{subfigure}%
\begin{subfigure}[t]{0.3\textwidth}
  \centering
  \includegraphics[width=\linewidth]{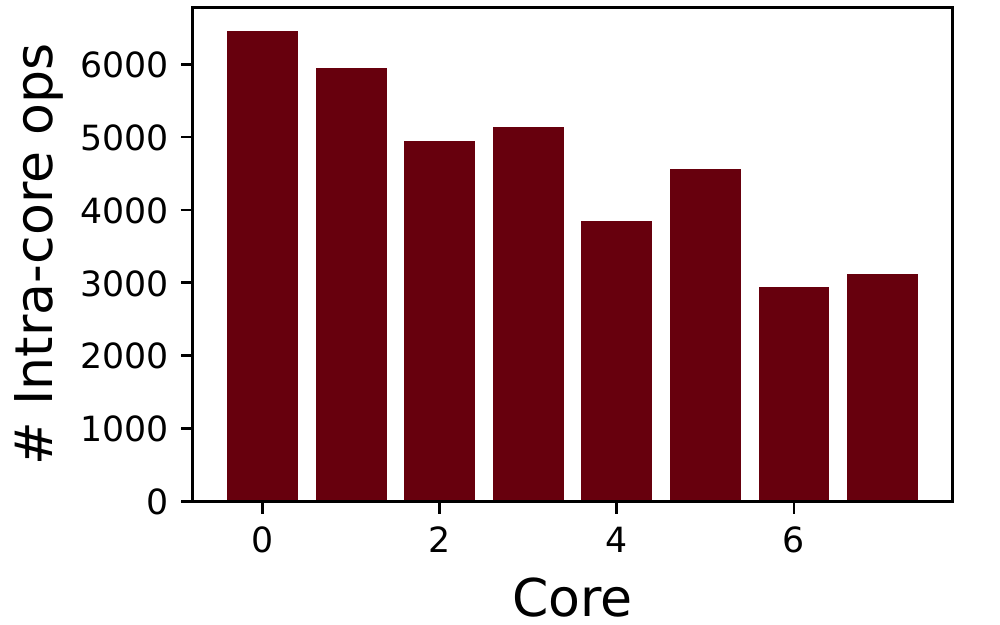}
  \vspace{-0.3cm}
  \caption{}
  \vspace{-0.1cm}
  \label{fig::traffic_heatmap_sub2}
\end{subfigure}
\begin{subfigure}[t]{0.38\textwidth}
  \centering
  \includegraphics[width=\linewidth]{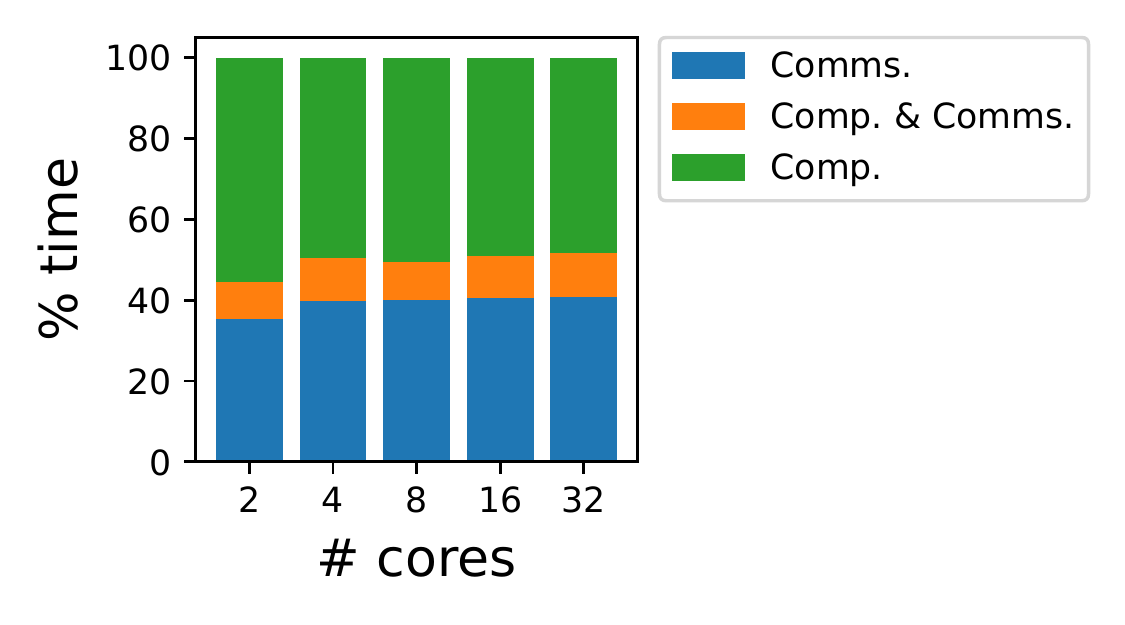}
  \vspace{-0.3cm}
  \caption{}
  \vspace{-0.1cm}
  \label{fig::traffic_heatmap_sub3}
\end{subfigure}
\vspace{-0.3cm}
\caption{\textbf{Qubit traffic in multi-core quantum architectures} \textbf{a)} Inter-core traffic, by pairs of communicating nodes. \textbf{b)} Intra-core operations (qubit gates) per core. \textbf{c)} Distribution of the execution time in computing and communicating in Cuccaro adder scaled in multi-core architectures. Observe the small amount of time in which computation and communications are allowed to be done in parallel.}
\vspace{-0.1cm}
\label{fig::traffic_heatmap}
\end{figure*}


Reviewing the main concepts on multi-core quantum computing and communication lets us set the stage for our work and quickly highlight the most important ``joints'' where the design analysis of multi-core quantum computing should be stressed. For a deeper look into quantum computing and communications, the interested reader may refer to \cite{national2019quantum, cacciapuoti2019quantum}.


\subsection{On qubits, gates and circuits}

The basic unit of quantum information is called a qubit. Similarly to its classical counterpart, a qubit can hold logical values, i.e. 0 and 1. However, due to \textit{quantum superposition}, its quantum state can also be a linear combination of both 0 and 1 states. Moreover, when two qubits are superposed through specific quantum gates, the quantum state becomes a combination of 00, 01, 10 and 11 states, in a process that can be extended to an arbitrary number of qubits. In other words, a quantum computer with $N$ superposed qubits is operating over $2^N$ states simultaneously, which provides an exponential increase in performance for certain applications.

When a qubit is measured, due to quantum mechanics' postulates, we obtain only a partial vision, and effectively \textit{destroy} the quantum state: the qubit \textit{collapses} into a deterministic value with certain probability. A derivative of this is the \textit{no-cloning} theorem, which translates into the impossibility of duplicating a quantum state. Two or more qubits can also be \textit{entangled}, i.e. whenever any of them is measured, all of them collapse into a definite state, with a non-zero correlation of the global result. For instance, two entangled qubits could be such that either both collapse to 0 or both collapse to 1. Along the present paper, we will differentiate \textit{physical qubits} (the material holders of the quantum data, whether they are movable, e.g. photons, or not, e.g. spin qubits) and \textit{virtual qubits} (the abstract quantum information which is operated on, swapped or teleported around the processor, measured, etc.).

These powerful properties allow for the \textit{quantum} leap in computing performance of quantum processors, but also introduce high complexity in the design and operation of them. In particular, qubits should be as isolated as possible from the outer environment, as the quantum state they hold is unstable and tends to decohere after a certain time: being unfeasible to copy or regenerate them, a corrupt qubit implies an irreversible information loss.

Quantum programs (also called quantum circuits) are described using quantum gates, i.e. logical gates applied on either one or two qubits, which modify their quantum state. In order to improve their isolation and minimize decoherence, qubits are operated and measured in-place. However, when two qubits are required to interact by means of a two-qubit gate, their quantum state needs to be moved to adjacent positions in the computer. This involves a resource and time overhead, that may be critical for the overall execution performance.

\subsection{A teamwork from quantum processors}

Multi-core quantum architectures represent an effort of utilizing currently existing limited-size quantum processors into a large structure by interconnecting them and making the whole platform work jointly (see Fig. \ref{fig::multi-core_computer}a). It may consist of dozens of quantum chips, containing hundreds to thousands qubits each, communicated both by a classical network (intended for signalling, message passing and measurement results communications) and a quantum backbone (to allow quantum state sharing).

Local operations on qubits are carried out as in single-core quantum processors. However, when a qubit from other node is required for an operation, a short-range quantum communication is performed. Although various types of techniques have been proposed, quantum teleportation is the most commonly assumed, due to its flexibility and robustness~\cite{cacciapuoti2020entanglement}.

Teleporting a qubit between two cores is a process involving a pair of entangled qubits (specifically, an EPR pair \cite{einstein1935can}), shared among the two communicating nodes, and a classical channel among them. To perform the teleportation, some basic operations involving the qubit to be transmitted and the entangled qubit at the transmission side are applied, followed by a measurement. The result (a binary value) is then sent via the classical channel. With that information, the receiver can reconstruct the original transmitted quantum state by applying some corrections. Note that by being measured, the original state of a qubit is lost and hence the no-cloning theorem is respected. Observe also that the quantum information transfer is done without physically moving the qubit holding it, but rather through the distribution of the EPR pairs and the classical movement of measurement information.

Therefore, quantum teleportation needs at least three things to work: a classical network communicating all processors, an EPR pair generator and a quantum network connecting it with all the nodes (see Fig. \ref{fig::multi-core_computer}b). Three different approaches may be used for the entanglement generation and distribution: \textit{i)} sharing an EPR pair generator among all nodes, using Spontaneous Parametric Down-Conversion, and thus having a star-like topology connecting the nodes with it; \textit{ii)} integrating on every core-to-core connection its own Bell State Measurement device, which is used to entangle photons coming from both transmitter and receiver nodes; \textit{iii)} using entanglement at source, i.e. generating at the transmitter node the pair of entangled particles, and sending out a photon to associate the remote node to the entanglement \cite{cacciapuoti2020entanglement}. For its low-resource requirements and simplicity, the first option is assumed for the rest of this paper.

\section{Using traffic analysis for performance analysis}

In classical multicore computers, the design of its internal Network-on-Chip (NoC) has become of extreme importance due to its impact on the performance of the entire processor. Since the design of any network requires an understanding of the traffic it needs to serve, considerable efforts have been spent over the years to characterize multicore systems and the applications that run on it.

Early works by Soteriou \emph{et al.} \cite{soteriou2006statistical} and Barrow \emph{et al.} \cite{barrow2009communication} analyzed a variety of multiprocessors between 16 and 32 cores running standard benchmark suites such as SPEC or PARSEC. In the former, the temporal burstiness, spatial hotspotness, and source-destination distance was studied, whereas in the latter, the focus was more on analyzing the memory sharing patterns leading to such traffic characteristics.

Subsequent studies pushed the analyses to larger systems up to 64 cores and delved into particular aspects, such as the time-varying characteristics of the traffic \cite{bogdan2011non}, which is often periodic as analyzed in \cite{gratz2010realistic}. This is due to the iterative nature of most algorithms running in multiprocessors, which further suggests that traffic is predictable. Further, the work in \cite{abadal2015multicast,abadal2016characterization} focused on multicast traffic only, demonstrating that such a subset of the workload is also bursty and predictable.

\begin{figure*}
    \centering
    \includegraphics[width=0.75\linewidth]{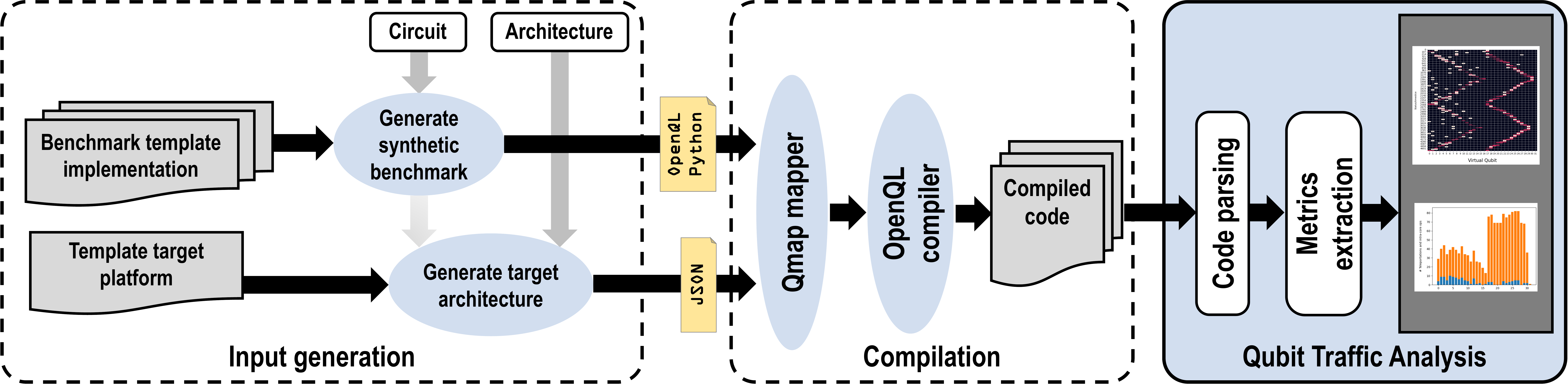}
    \vspace{-0.1cm}
    \caption{Flow diagram of the qubit traffic analysis tool}
    \vspace{-0.1cm}
    \label{fig::traffic-tool-flow}
\end{figure*}

These workload characterization studies had several impacts on the NoC field. In particular, they allowed to \textit{i)} study aspects such as the correlation between particular traffic characteristics and on-chip network congestion \cite{gratz2010realistic}, \textit{ii)} create synthetic traffic generators better reflecting real workloads for the evaluation of NoC designs \cite{soteriou2006statistical,bogdan2011non,abadal2016characterization}, and eventually, \textit{iii)} guide the design of improved topologies, routing policies, or congestion control mechanisms at the chip scale.


A pertinent question is then whether a similar approach can be used to characterize the workload of quantum processors. Before answering that question, though, it is important to see the main differences between both worlds. In classical systems, most of the traffic is an implicit consequence of the memory accesses produced by a multithreaded application and, hence, very hard to infer from compiled code. In quantum algorithms, on the other hand, communication primitives are explicit in the compiled code so that the traffic becomes not predictable, but rather known beforehand. Another difference is that due to the no-cloning theorem, it is hard to envisage the need for multicast communication at least resulting directly from the need to move the quantum state of qubits. Other than that, the metrics used in classical computing or the insight gained through analysis of its workloads, such as the iterative nature of communication, can be still useful in the quantum world.



\section{A qubit traffic analysis software tool}

\begin{figure*}
\centering
\begin{subfigure}[t]{0.495\textwidth}
  \centering
  \includegraphics[width=\linewidth]{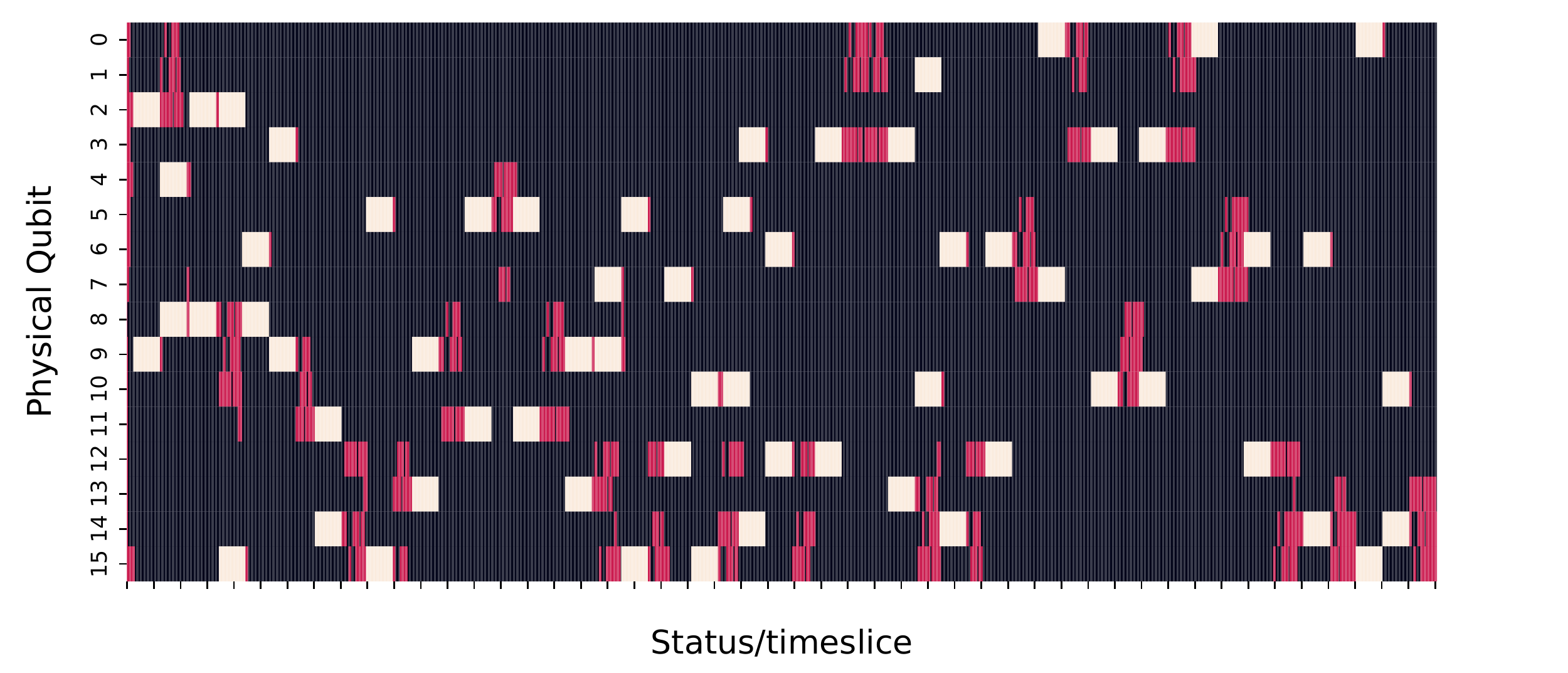}
  \vspace{-0.3cm}
  \caption{}
  \vspace{-0.1cm}
  \label{fig::grovers_analysis_sub1}
\end{subfigure}%
\begin{subfigure}[t]{0.495\textwidth}
  \centering
  \includegraphics[width=\linewidth]{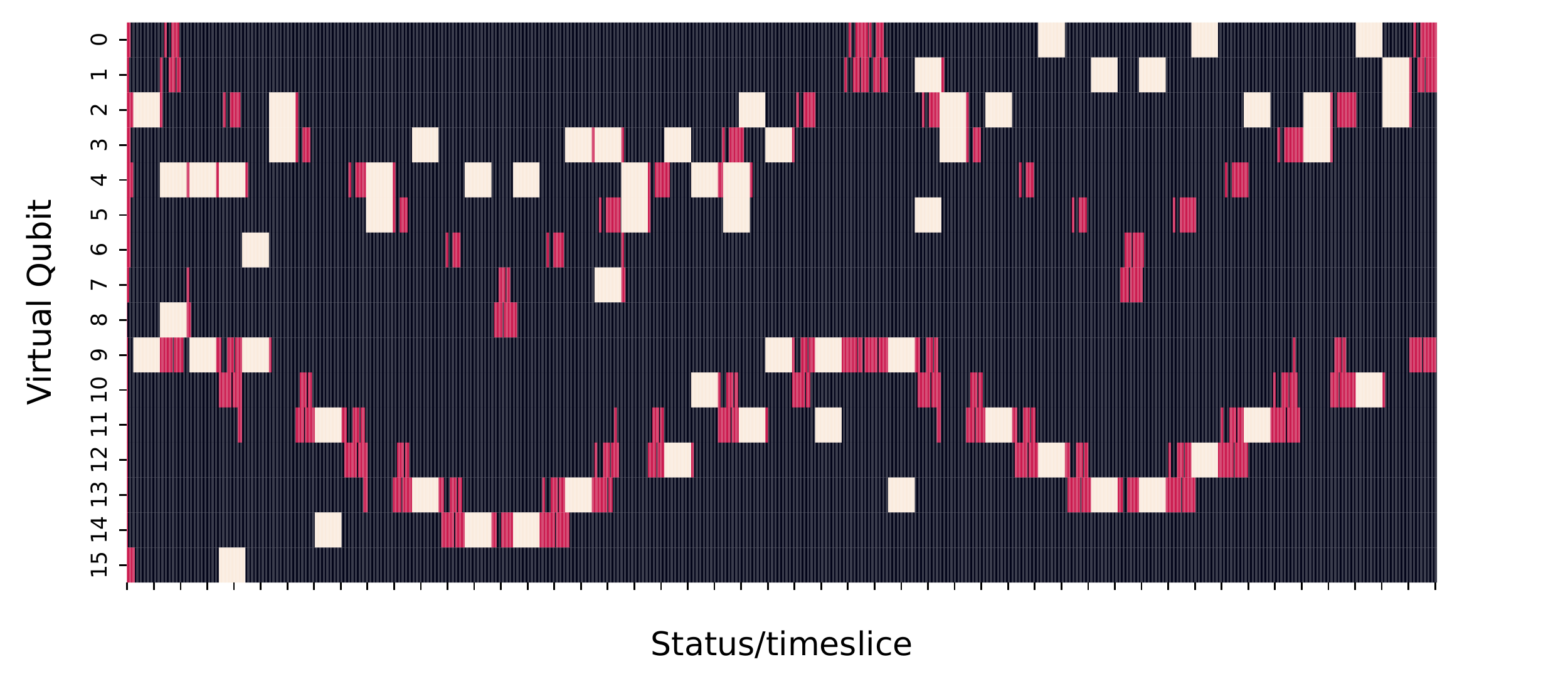}
  \vspace{-0.3cm}
  \caption{}
  \vspace{-0.1cm}
  \label{fig::grovers_analysis_sub2}
\end{subfigure}
\caption{\textbf{Execution trace of Grover's main routine for 16 qubits} \textbf{a)} on physical qubits and \textbf{b)} on virtual qubits. Computation, communication, and idling times are represented in red, white, and black colors respectively.}
\label{fig::grovers_analysis}
\end{figure*}

Building a tool for analyzing traffic during the execution of quantum circuits in multi-core quantum architectures calls for firstly understanding where the traffic comes from and what are the main sources and stakeholders of traffic during the whole process.

In a generic multi-core quantum platform qubits are constantly moving around. Indeed, as stated in Section 2, whenever two distant qubits (even if they are on the same core) are to be operated by means of a two-qubit gate, they must be moved to adjacent positions. This implies that quantum circuits involve constant qubit traffic, both in and between cores. See Fig. \ref{fig::multi-core_computer}c for a simple example on how qubit traffic is generated by both intra-core topology constraints and inter-core operations. In Figs. \ref{fig::traffic_heatmap_sub1} and \ref{fig::traffic_heatmap_sub2}, a simulated example based on a real quantum circuit is shown: specifically, the aggregated node-to-node and intra-node traffic of a sample execution of the QFT circuit of 128 qubits on a 8-core platform with 16 qubit per core. Observe the high total count of teleportations among cores, and the existence of some hotspots, attracting most of the communication and computation (cores 0 and 1). 

In addition, inter-core communication is slower than intra-core communication operations: latencies are from 5$\times$ to 100$\times$ longer \cite{baker2020time, monroe2014large}. This, together with the generally high dependency between gates, leads to almost idle execution intervals following high intensity ones. In the same example as before, see in Fig. \ref{fig::traffic_heatmap_sub3} the time distribution of computation (execution of qubit gates) and communication (teleportation operations) when scaling Cuccaro adder circuit in multi-core architectures. Most of the time, the dependencies present in the circuit make the processor to idle while waiting for teleportations to end (in the example, only about 10\% of the execution there are simultaneous computation and communication operations).

Therefore, knowing that all this communications overhead impacts on the reliability of the computation, we would desire to minimize these movements and equalize the traffic. Let us quickly review the three main stakeholders involved in traffic generation and control:
\begin{itemize}
    \item \textbf{The quantum circuit}. The number and distribution of two-qubit gates will impact on the qubit traffic during execution.
    \item \textbf{The processor's topology}. A scarcely connected processor leads to a higher communications overhead when mapping two-qubit gates into the circuit. Moreover, in a multi-core scenario, the lower the ratio of number of qubits per core to number of cores, the higher the need for costly inter-core qubit communications.
    \item \textbf{The compiler algorithm (mapper and scheduler)}. When compiling the quantum circuit into a physical platform, optimizations can be applied to allow for minimizing the traffic overhead.
\end{itemize}

These would be the \textit{deterministic} sources of traffic. There could be also some impact coming from communication errors during the execution, which could lead to unbounded communication latencies. However, we have decided not to include them into this work's analysis, so as to study them separately in future work including fully-fledged simulations.

Therefore, we have developed a software tool that, given a quantum circuit and a target many-core quantum platform, allows us to extract the qubit traffic by tracing all qubits along the execution and registering all the gates they participate in and the moves they are involved in.

The process, graphically explained in Fig. \ref{fig::traffic-tool-flow}, consists on the following steps: \textit{i)} generation of the quantum circuit with the corresponding qubit input length, \textit{ii)} compilation of the quantum circuit on the target platform, always having the same number of physical qubits as the \textit{width} (qubits involved) of the quantum circuit and the required number of cores, and \textit{iii)} parsing of the resulting cQASM code in order to obtain the trace of each of the qubits. That information may be used for analysis purposes studying traffic burstiness, hotspots and other related metrics.

For the compilation process, we have used OpenQL \cite{khammassi2020openql} and a modified version of the Qmap mapper \cite{lao2019timing} embedded in it, extended to the multi-core case following proposal from Baker \textit{et al.}~\cite{baker2020time}. All the software has been programmed using Python 3.8.

\subsection{Looking at a quantum circuit in a different way}

\begin{figure*}
\centering
\begin{subfigure}[t]{0.2\textwidth}
  \centering
  \includegraphics[width=\linewidth]{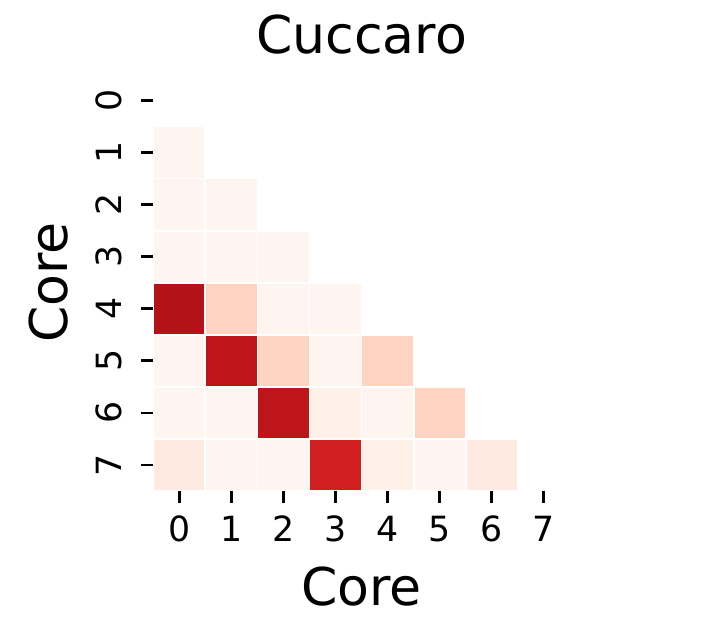}
  \vspace{-0.3cm}
  \vspace{-0.1cm}
  \label{fig::heat_cuccaro}
\end{subfigure}%
\begin{subfigure}[t]{0.2\textwidth}
  \centering
  \includegraphics[width=\linewidth]{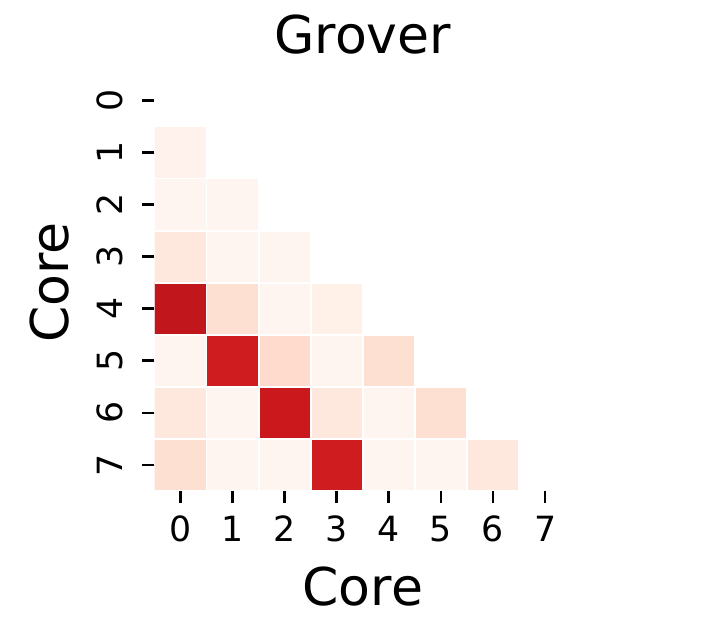}
  \vspace{-0.3cm}
  \vspace{-0.1cm}
  \label{fig::heat_grover}
\end{subfigure}%
\begin{subfigure}[t]{0.2\textwidth}
  \centering
  \includegraphics[width=\linewidth]{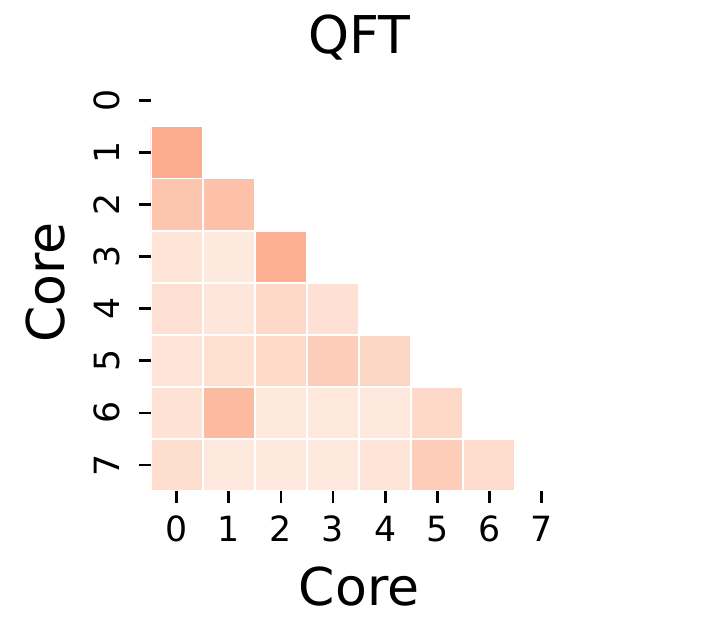}
  \vspace{-0.3cm}
  \vspace{-0.1cm}
  \label{fig::heat_qft}
\end{subfigure}%
\begin{subfigure}[t]{0.2\textwidth}
  \centering
  \includegraphics[width=\linewidth]{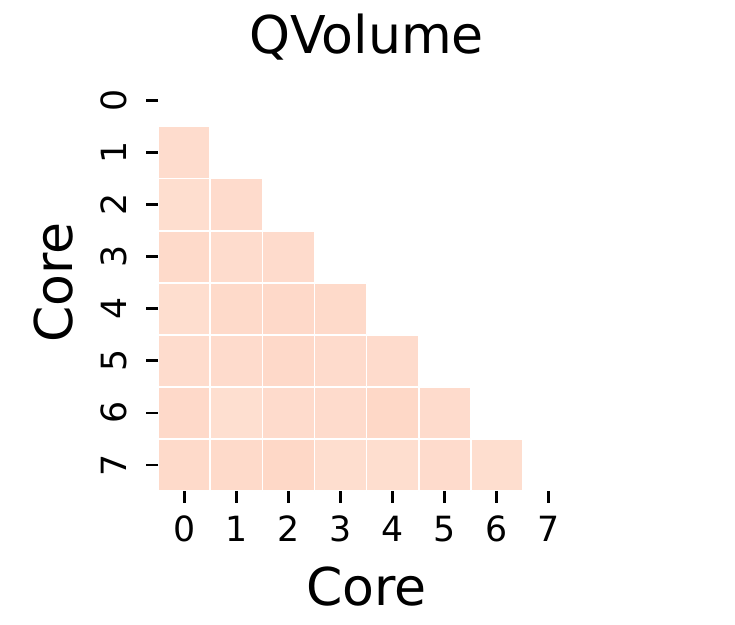}
  \vspace{-0.3cm}
  \vspace{-0.1cm}
  \label{fig::heat_qv}
\end{subfigure}%
\begin{subfigure}[t]{0.2\textwidth}
  \centering
  \includegraphics[width=\linewidth]{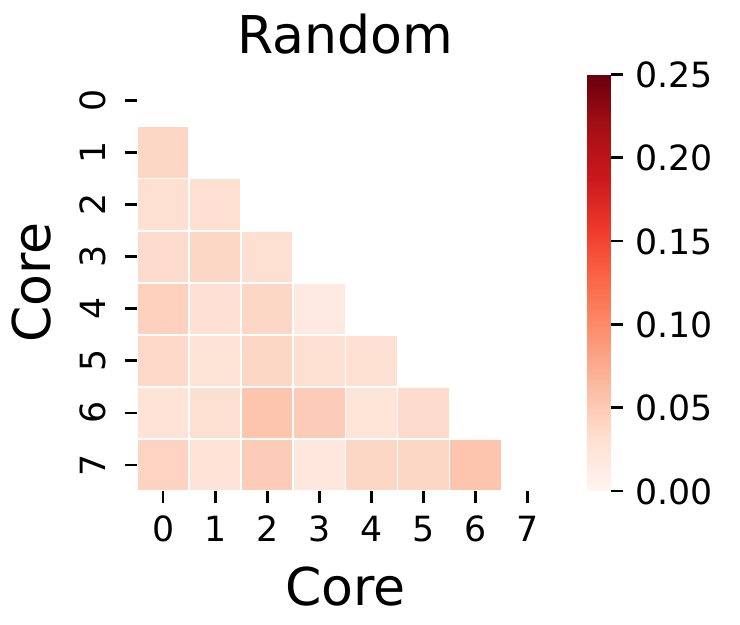}
  \vspace{-0.3cm}
  \vspace{-0.1cm}
  \label{fig::heat_rnd}
\end{subfigure}%
\caption{Inter-core traffic as the ratio of the number of teleportations for every pair of communicating nodes over total teleportations in all benchmarks assuming 8 cores with 16 qubits per core.}
\label{fig::heat_algorithms}
\end{figure*}

Seeing the execution from this perspective gives us some insights on the performance of both the algorithm itself and the mapper, as we may easily observe how strong dependencies that block execution are dealt, the efficiency of the overall execution, idling periods, distribution of the job among the available cores and physical qubits, as well as analyzing the ``life'' of any virtual qubit along the execution.

Let us analyze a single example with our tool to see how these things may be observed and analyzed. We will use a small example for better visualization: the Grover's main routine for 16 qubits run in a 4-core platform (4 qubits per core). In Fig. \ref{fig::grovers_analysis} we present some graphical views of the execution to help us in the analysis. See in Fig. \ref{fig::grovers_analysis_sub1} the distribution of gate executions (in red) and teleportations (in white) versus the idling times (in black). The x axis is the timeline of the execution (the beginning at the top) and the y axis corresponds to the physical qubits in the system (ordered by core). The amount of dependencies among operations involves a high execution inefficiency, as most of the time qubits are idle, waiting for a single operation (either a quantum gate or a teleportation) to finish. The qubits in the bottom-most core are clearly accumulating the result and do the most computation and communication. In Fig. \ref{fig::grovers_analysis_sub2}, the y axis is now the virtual qubits, hence we see the logical operation of the algorithm more clearly. Now we can observe the dependencies better, and can differentiate virtual qubits that are expected to last with a coherent state for almost all the computation (e.g. qubit 9), whereas others are almost of no use (e.g. qubit 15).

These and other conclusions can be extracted using our tool, making it easier to give design guidelines for the algorithm, the architecture designer and the compiler engineer.

\section{Experimental results}
In the previous section we have focused on analyzing a single execution, seeing that several straightforward conclusions can be extracted. However, by aggregating these qualitative observations into numerical metrics and, in the future, using design space exploration techniques, we can obtain even more interesting insights.

\subsection{Simulation set up and architectural space}

In this paper, we introduce our tool and present some of the results obtained with it. Here we have restricted the exploration to a smaller set of examples, and will fully develop it on a larger upcoming paper. In particular, we have used multi-core architectures with the following fixed characteristics: the cores are interconnected via a classical network for exchanging classical messages and measurements. All nodes are connected via optical channels to an EPR pair generator instrumental for qubit teleportation. The teleportation operation is assumed by the compiler as deterministically time-bound (set to 1000 ns, around 4 times more than a SWAP gate, in order to stress the system \cite{rodrigo2021scaling}), and always performed as a SWAP operation, i.e. the two qubits involved are swapped after the teleportation operation. The connectivity inside every core is full, i.e. any qubit can perform a 2-qubit gate with any other in the same core. This is done to ``isolate'' in the analysis the inter-core communication from the intra-core computation, which does not involve extra-SWAPs. The rest of gate and qubit parameters are taken from typical superconducting flux qubits.

The exploration has been done by analyzing various algorithms (both real applications and random benchmarks, see below) and different platform configurations, varying the number of cores and number of qubits per core. In all cases, the circuits compiled in a given platform occupy all physical qubits available.


\subsection{The selected algorithms}

As benchmarks for assessment of communication overhead for multi-core architectures and their scalability, we opted for several algorithms that have potential to show computational advantage when run on quantum in comparison to classical computers, such as Quantum Fourier Transform (QFT), Grover's search algorithm and Cuccaro Adder. These algorithms, however, have a specifically defined structure which makes them scale with number of qubits in steady, sometimes even linear way (Grover's), in terms of their parameters like number of gates or two-qubit gate percentage. For that reason we additionally used randomly generated algorithms as well as quantum volume circuits \cite{cross2018validating}, where we could have more influence on their parameters for any size of the circuit, and therefore probe our architecture in a worst-case scenario. The random algorithms we used were generated with uniformly chosen gates from a limited gate set with uniform distribution of those gates among qubits. Quantum volume circuits are used in general for probing even single-core architectures, as they are the most complex version of synthetic circuit with the highest two-qubit gate density (forces all qubits to be engaged in a two-qubit gate in each circuit layer).

\subsection{Explorations}

Following literature on traffic analysis for multi-core scenarios (see Section 3), we have performed our exploration on the selected benchmarks in a three-phase fashion: first, studying the temporal distribution of the quantum data trasnfers; then, focusing on their spatial distribution, and finally, summarizing both analysis in a spatio-temporal joint exploration.

\begin{figure}
    \centering
    \includegraphics[width=\linewidth]{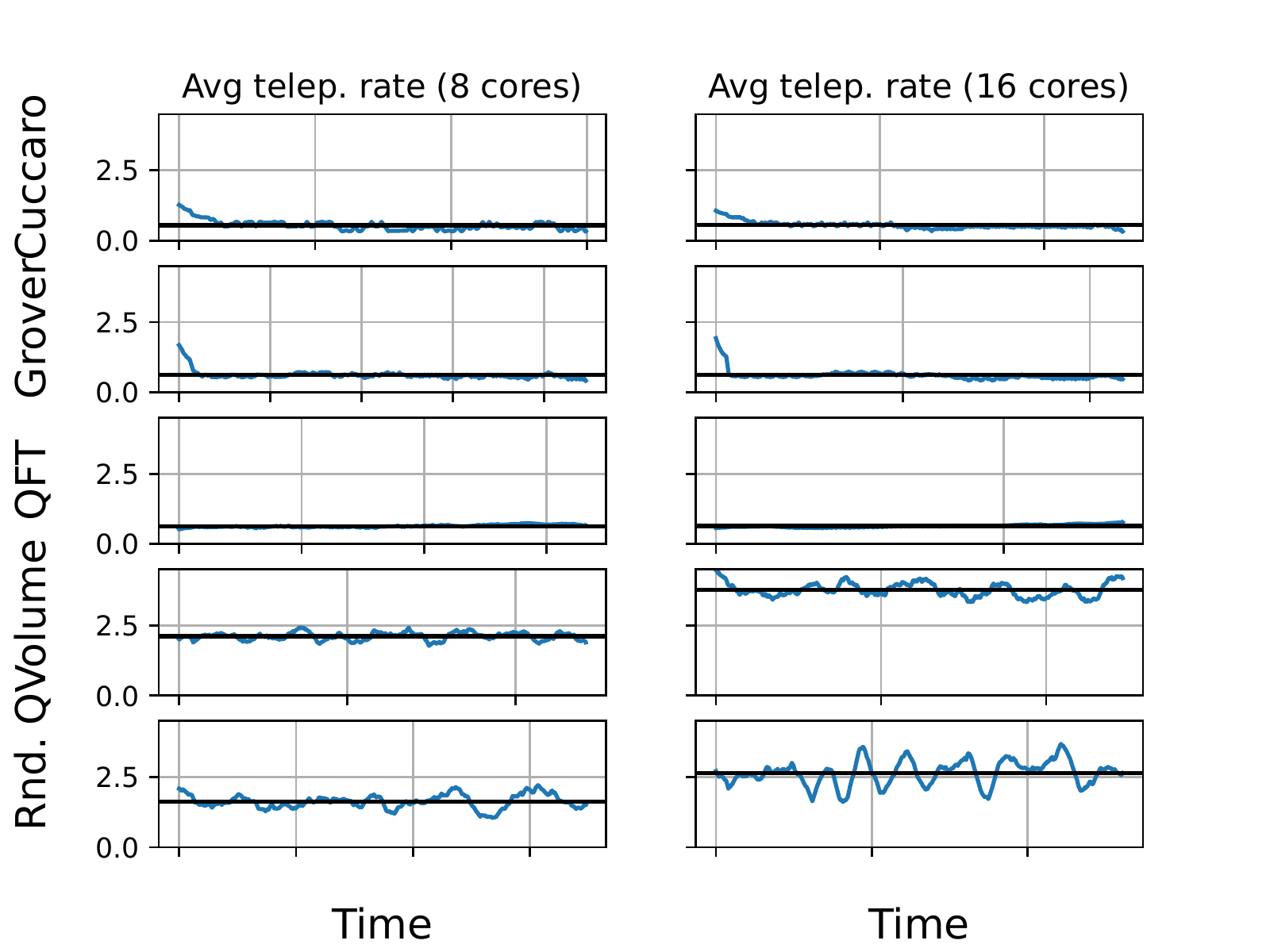}
    \vspace{-0.1cm}
    \caption{Average number of teleportations per timeslice in all benchmarks assuming 8 qubits per core and either 8 or 16 cores.}
    \vspace{-0.1cm}
    \label{fig::moving_mean}
\end{figure}

For the temporal traffic distribution, we have studied the inter-core communication trends for the different algorithms. In Fig. \ref{fig::moving_mean}, the moving average of the number of teleportations per timeslice in every circuit, together with the overall mean, is plotted for all algorithms. Two different cases are studied (8 and 16 cores, both with 8 qubits per core). Observe that both Cuccaro and Grover suffer from a high inter-core data transfer burst at the start, which may easily stress the system and cause a bottleneck on loaded or poorly-connected architectures. Both of them, together with QFT, have a quite low average number of teleportation (around 1), which is mostly related to the dependencies among operations in the code, forcing an almost linear, non-parallel, execution (as already observed in Fig. \ref{fig::traffic_heatmap}). Random and Quantum volume cases are good to stress the system, as they have more relaxed dependencies and allow for a higher teleportation rate. This communications requirements scale with the number of cores: this does not seem to be the case for Grover, Cuccaro and QFT, which may facilitate scaling on large multi-core architectures.

For the spatial traffic analysis, i.e. how evenly is the overall traffic distributed among the cores, we have focused on whether the compile circuit creates hotspots (cores attracting most communications). Hotspotness may be a natural consequence of most circuits, that e.g. concentrate the result on a given variable, but it results in network congestion. In Fig. \ref{fig::heat_algorithms}, the inter-core traffic is presented for all benchmarks, for the 8 cores, 16 qubits per core case. The random and Quantum Volume cases are quite uniform, as expected, while that is also the case for QFT. Being a core part of some key quantum algorithms, avoiding network bottlenecks in QFT on multi-core architectures is relevant for their overall computational performance. Still, some minor hotspots (cores 0 and 1) can be detected, and probably further optimizations in the compiler could fix that. Grover and Cuccaro present a very similar behavior, which most probably has to do with the initial burst and flaws in the qubit mapping.

\begin{figure}
    \centering
    \includegraphics[width=\linewidth]{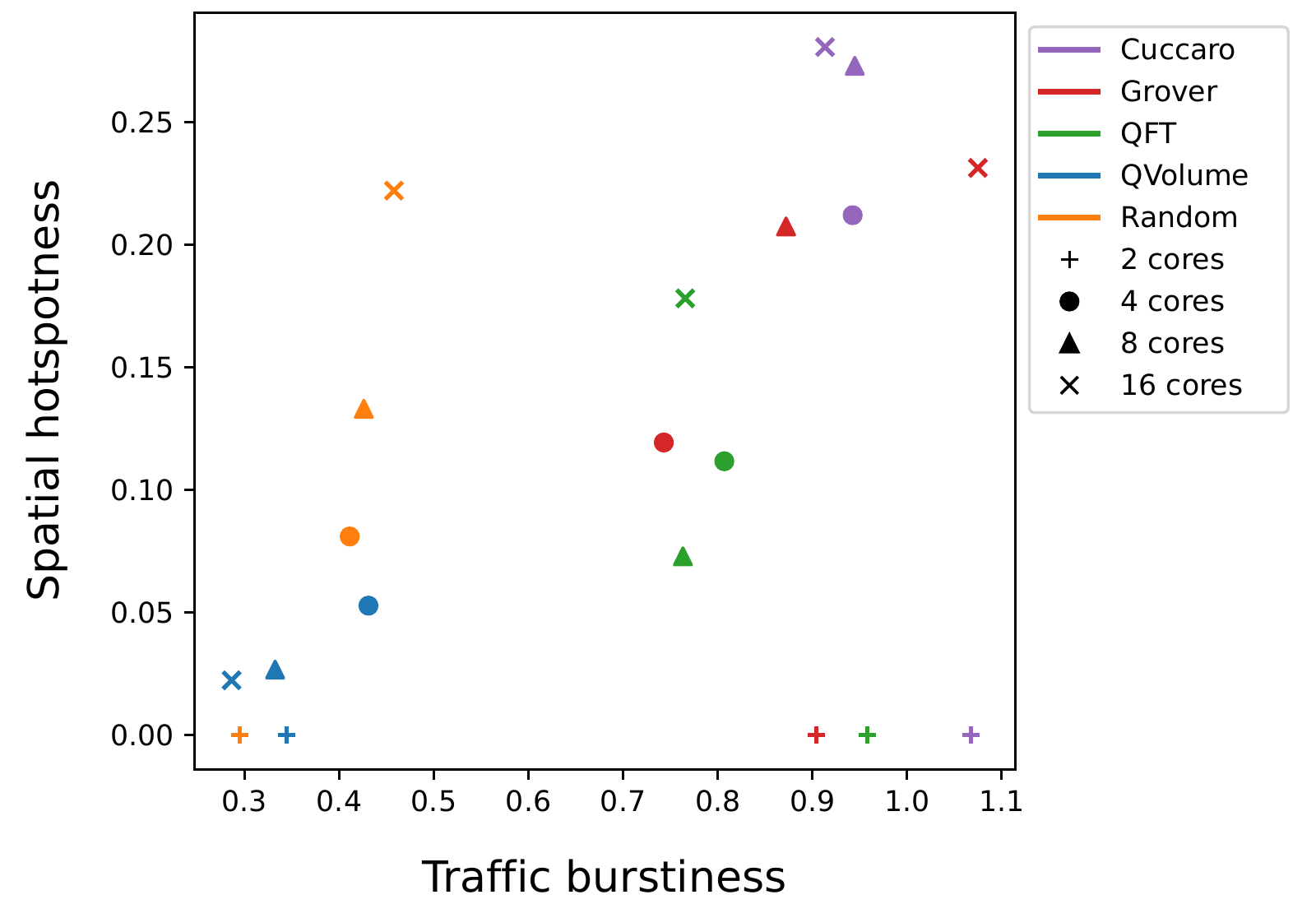}
    \vspace{-0.1cm}
    \caption{Summary of burstiness and hotspotness of all the evaluated benchmarks and core counts assuming 8 qubits per core.}
    \vspace{-0.1cm}
    \label{fig::spatio-temporal}
\end{figure}

Finally, a joint spatio-temporal analysis is performed, using results as plotted in Fig. \ref{fig::spatio-temporal}. A wider exploration is performed, for a core count ranging from 2 to 16 cores (8 qubits per core). We have used covariance (standard deviation $\sigma$ over the mean) of both spatial and temporal traffic. For the spatial hotspotness we have used the number of teleportations per core over the whole execution, and for the temporal burstiness, we have used the number of teleportations per timeslice. Observe that there are two differentiate regions: random and Quantum Volume have low burstiness, while the rest are on the high burstiness end. Burstiness results in network inefficiencies due to unexpected bottlenecks and calls for overdimensioning the network capacity. See also that in general, the spatial hotspotness is specially high for Cuccaro, and that Grover scales quickly while QFT does it in a more controlled way.

    \vspace{-0.4cm}
\section{Conclusions}
\label{sec:conclusion}

In this paper we have substantiated the interest of qubit traffic analysis for efficient multi-core quantum architectures, and presented a tool for carrying out this characterization. We have showcased, with some first explorations, how traffic metrics may help in quantum algorithms classification, optimization of compilers for multi-core quantum architectures, and highlight the communications requirement for a given application and target architecture.

We plan to use this tool to do further explorations with larger sets of benchmarks and range of target architectures, as well as complementing this analysis with fully-fledged simulations that may shed the light on online quantum network management for error mitigation. Also, it is worth exploring which structural parameters in quantum circuits are the reason behind most inefficiencies found (data transfer bursts, hotspots, code dependencies...). This in-depth analysis might help us improve our inter- and intra-core communication strategy and later on give us the guidelines for multi-core device design that is more compatible with specific type of algorithms.


\end{document}